\documentclass[]{aastex701}

\begin{document}

\title{LAP1-B is the First Observed System Consistent with Theoretical Predictions for Population III Stars}


\author[]{Eli Visbal}
\affiliation{Department of Physics and Astronomy and Ritter Astrophysical Research Center, University of Toledo, 2801 W. Bancroft Street, Toledo,
OH 43606, USA}
\email[show]{Elijah.Visbal@utoledo.edu}  

\author[]{Ryan Hazlett} 
\affiliation{Department of Physics and Astronomy and Ritter Astrophysical Research Center, University of Toledo, 2801 W. Bancroft Street, Toledo,
OH 43606, USA}
\email{Ryan.Hazlett@rockets.utoledo.edu}

\author[]{Greg L. Bryan}
\affiliation{Department of Astronomy, Columbia University, New York, NY 10027, USA}
\email{gb2141@columbia.edu}


\begin{abstract}
Recently, Nakajima et al.~(2025) presented James Webb Space Telescope observations of the $z=6.6$ Population III (Pop III) candidate LAP1-B, which is gravitationally lensed by galaxy cluster MACS J0416. We argue that this is the first object to agree with three key theoretical predictions for Pop III stars: (1) formation in extremely low-metallicity halos with virial temperatures ranging from $T_{\rm vir}\sim 10^3-10^4~{\rm K}$, (2) a top-heavy initial mass function, and (3) formation of low-mass clusters with ${\sim}{\rm a ~few}\times 1000~M_\odot$ in massive Pop III stars. LAP1-B is consistent with recently formed Pop III stars hosted in a $\sim 5\times 10^7~M_\odot$ dark matter halo, some of which have enriched their surrounding gas either with supernovae or stellar winds. We use the semi-analytic model of Visbal et al. (2020) to predict the abundance of Pop III clusters observable at the high magnification provided by the foreground galaxy cluster MACS J0416. Using fiducial parameters unmodified from previous work, we expect about one observable  Pop III galaxy similar to LAP1-B in the range $z=6-7$. At earlier times, the intrinsic abundance is higher, but Pop III systems would not have been detected because of their increased luminosity distance and lower mass dark matter halos, which would host fewer stars. Thus, we find that LAP1-B was found at the redshift theoretically expected, given current observable limits, despite the fact that most Pop III systems form much earlier. 
\end{abstract}

\keywords{\uat{Population III Stars}{1285} --- \uat{Cosmology}{343}}


\section{Introduction}
\label{sec:intro}
Understanding the formation and properties of the first stars in the Universe is currently an exciting frontier in astrophysics and cosmology. Up to this point, there have been no unambiguous direct detections of Population III (Pop III) stars, defined by their extremely low metallicities. However, substantial theoretical work over the past several decades has led to a consensus regarding several key predictions  \citep[for a recent review see][]{2023ARA&A..61...65K}.

First, \emph{Pop III stars form in extremely low metallicity halos with virial temperatures ranging from $T_{\rm vir} \approx 10^3- 10^4~{\rm K}$}, corresponding to virial masses of $M_{\rm vir}=10^6 \left (\frac{1+z}{11} \right )^{-3/2} - 3 \times 10^7 \left (\frac{1+z}{11} \right )^{-3/2}$  ({\bf ``Prediction 1''}).
The earliest Pop III stars are expected to form at $z\gtrsim 20$ in the smallest halos in this range \citep{1996ApJ...464..523H, 1999ApJ...527L...5B, 2002Sci...295...93A, 2003ApJ...592..645Y}. These halos are the first to create molecular hydrogen and excite its rovobrational transitions leading to radiative cooling and star formation. 
However, several physical processes can prevent star formation in low-mass dark matter halos.
Radiation from these stars in the Lyman-Werner (LW) band photodissociates molecular hydrogen and can delay star formation in halos up to $T_{\rm vir}\approx 10^4~\rm {K}$, at which point atomic hydrogen cooling becomes important \citep{1997ApJ...476..458H, 2001ApJ...548..509M, 2008ApJ...673...14O,2014MNRAS.445..107V}. 
In addition, ionizing radiation photoheats gas, suppressing star formation in halos up to $T_{\rm vir}\approx 3\times 10^4~\rm {K}$ \citep{1998MNRAS.296...44G, 1996ApJ...465..608T, 2004MNRAS.348..753S, 2004ApJ...601..666D, 2006MNRAS.371..401H, 2014MNRAS.444..503N}.
The baryon-dark matter streaming velocity additionally prevents star formation in low-mass, high-redshift halos \citep{2010PhRvD..82h3520T,2012MNRAS.424.1335F}.
On the high virial temperature (or high halo mass) end, prompt metal production by the first generation of stars results in a transition to Pop II stars. However, inhomogeneous metal mixing results in Pop III star formation continuing until $z\sim 6$ \citep[e.g., see Figure 2 in ][]{2023ARA&A..61...65K}, when the abundance of such systems drops rapidly.

The second key theoretical prediction is that \emph{Pop III stars have an initial mass function (IMF) which approximately follows $N_*(M_*)d\log(M_*)\propto {\it const.}$} \citep[see Fig.~6 in][]{2023ARA&A..61...65K} ({\bf ``Prediction 2''}). The exact form of the IMF remains uncertain, but many groups have found roughly similar top-heavy distributions with ranges from ${\sim}10-1000~M_\odot$ \citep[e.g., ][]{2015MNRAS.448..568H}.

The final theoretical prediction we highlight is that \emph{Pop III stars form in low-mass clusters of up to ${\sim}{\rm few}\times 1000~M_\odot$ total in massive stars} ({\bf ``Prediction 3''}). This is a result of fragmentation, rapid radiative feedback, and rapid metal pollution, which have been observed in a variety of simulations \citep{2016ApJ...823..140X, 2019ApJ...882..178K, 2023MNRAS.524..351K, 2025ApJ...980...41B}. We emphasize that throughout this Letter we focus on halos where star formation is occurring for the first time and thus is composed entirely of Pop III. We refer to this as ``classical'' Pop III star formation. We do not consider the case of inhomogeneous metals within galaxies and mixtures of Pop III and metal-enriched stars \citep[e.g.,~][]{2022ApJ...935..174S}.

\emph{JWST} has detected several Pop III candidates from the epoch of reionization (EoR) via their colors \citep{glimpse} and emission lines \citep{2024ApJ...967L..42W, 2024A&A...687A..67M}. 
Recently, \cite{LAP1B} presented a spectrum of LAP1-B, which is a highly magnified ($\mu\sim 100$) source at $z=6.6$ \citep{2023A&A...678A.173V, LAP1B}.  Based on their measurements of H, C, and O lines, and the absence of detectable continuum emission, \cite{LAP1B} conclude that LAP1-B is most likely a cluster of $\lesssim 2700~M_\odot$ Pop III stars. The metals are suggested to have been recently provided by Pop III supernovae from the same cluster of stars. Additionally, \cite{LAP1B} estimate the total gas mass of LAP1-B within 20 pc to be ${\sim}4\times 10^5~M_\odot$ based on the Kennicutt-Schmidt relation (where the star formation rate is estimated from the H$\alpha$ line). This is consistent with an upper bound on the gas mass they calculate from the oxygen yield of the stars. 

There have been a number of previous claims for the detection of classical Pop III galaxies. In this Letter, we argue that \emph{LAP1-B is the first Pop III candidate that is consistent with all three of the theoretical predictions described above.} We use the semi-analytic model of \cite{2020ApJ...897...95V} to estimate the abundance of highly magnified Pop III sources similar to LAP1-B that were expected to be found in the MACS J0416 galaxy cluster field given the JWST flux limit. We find that observing ${\sim}1$ Pop III galaxy at $z {\sim}6.5$ is likely, and that the detectability of similar objects drops off rapidly at higher redshifts. This abundance estimate provides additional strong theoretical support that LAP1-B is hosted by a low virial temperature, atomic cooling halo (Prediction 1) and shows that $z{\sim}6.5$ is the most natural redshift for a first detection of Pop III stars. Additionally, LAP1-B's line emission is consistent with a recently formed starburst of several thousand massive Pop III stars (Prediction 3) from a top-heavy IMF that has started to pollute its surrounding gas through supernovae or stellar winds (Prediction 2). 

The remainder of this Letter is structured as follows. In Section \ref{sec:properties}, we discuss the physical properties of Pop III galaxies expected at $z\sim6.5$ based on previous theoretical predictions and compare them to the observations of \cite{LAP1B}. In Section \ref{sec:abundance}, we compute the abundance of Pop III galaxies expected to be observable in a cluster similar to MACS J0416. We discuss these results and compare with other previous Pop III candidates before concluding in Section \ref{sec:conclusions}. Throughout this work, we assume a $\Lambda {\rm CDM}$ cosmology with parameters from \cite{2014A&A...571A..16P}: $\Omega_{\rm m} = 0.32$, $\Omega_{\Lambda} = 0.68$, $\Omega_{\rm b} = 0.049$, $h=0.67$, $\sigma_8=0.83$, and $n_{\rm s} = 0.96$. These values were chosen to match the set of N-body simulations used in our semi-analytic model for our abundance estimate.

\section{Physical Properties of LAP1-B} 
\label{sec:properties}
\subsection{Mass of Pop III Stars}
\label{sec:popIII_mass}
We begin by discussing the H$\alpha$ flux from LAP1-B, which is the highest signal-to-noise line detected in \cite{LAP1B} (${\rm S/N{\sim}8}$). H$\alpha$ is a recombination line, and thus its flux can be related to the rate of H-ionizing photons produced by massive stars. We note that the observed H$\alpha$/H$\beta$ ratio indicates negligible dust extinction. Following \cite{2002A&A...382...28S}, the line luminosity is then given by $L_{\rm H\alpha} = c_{\rm H\alpha}(1-f_{\rm esc})Q_{\rm tot}(H)$, where $c_{\rm H\alpha}= 1.21 \times 10^{-12}~{\rm erg}$, $f_{\rm esc}$ is the escape fraction of the ionizing photons and $Q_{\rm tot}(H)$ is the rate of H-ionizing photons produced by the stars. 
As discussed above, hydrodynamical simulations suggest that the Pop III IMF is predicted to be logarithmically flat and likely over a massive range (Prediction 2). 
We note that for this IMF shape, the majority of the contributions to the stellar mass and the ionizing photons come from the most massive stars near the upper mass limit of the IMF. Thus, roughly speaking, for a logarithmically flat IMF, we can think of the properties of a Pop III cluster as being set by a characteristic stellar mass scale. 

\begin{figure*}
    \centering
    \includegraphics[width=10cm]{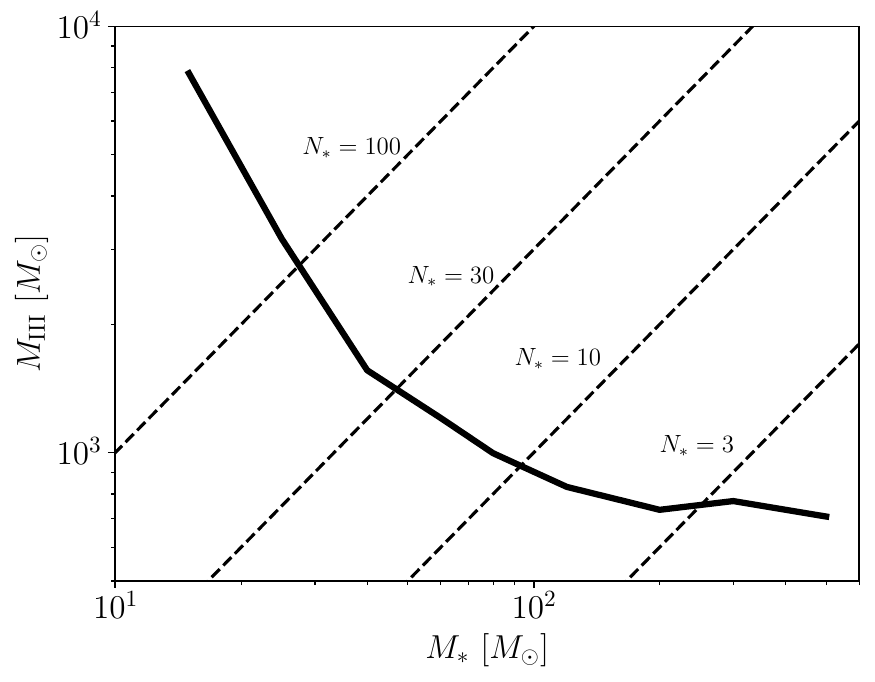}
    \caption{The total stellar mass of Pop III stars required to generate the H$\alpha$ emission observed in LAP1-B as a function of the mass of each individual star (computed from Eq.~\ref{eqn:mass} assuming all stars are of the same mass). This value is calculated assuming fractional numbers of stars are allowed which is unphysical, but shows that the ionizing efficiency saturates at stellar masses of ${\sim}200~M_\odot$. Diagonal dashed lines indicate constant numbers of stars.
    \label{fig:stellar_mass}  }
\end{figure*}

Assuming all of the massive Pop III stars in LAP1-B are of a single mass, the observed H$\alpha$ flux of $2\times 10^{-19}~{\rm erg~s^{-1}~cm^{-2}}$ from $z=6.6$, would require a total stellar mass of  
\begin{equation}
\label{eqn:mass}
M_{\rm III} = 1500 ~ M_\odot \left( \frac{100}{\mu} \right ) \left( \frac{0.9}{1-f_{\rm esc}} \right )\left( \frac{M_*}{40~M_\odot} \right )\left( \frac{2.469\times 10^{49} ~{\rm s^{-1}}}{Q(H)} \right ),
\end{equation}
where $\mu$ is the gravitational lensing magnification factor for LAP1-B, $M_*$ is the mass of individual Pop III stars, $Q(H)$ is the lifetime-averaged rate of H-ionizing photon production for a single star, and $M_{\rm III}$ is the total mass of Pop III stars currently emitting radiation. 
Note that $Q(H)$ depends on the assumed stellar mass. 
Thus, taking the relevant values for non-rotating stars in \cite{2002A&A...382...28S} we find that LAP1-B could be composed of approximately $38$ Pop III $40~M_\odot$ stars. Higher-mass stars have an increased ionizing efficiency, so less total stellar mass would be required. We illustrate this in Figure \ref{fig:stellar_mass}. We note that ${\sim}1000~M_\odot$ of massive Pop III stars is in agreement with Prediction 3. The emission-line ratios from \cite{LAP1B} are indicative of massive Pop III stars (see their Fig.~3), which is consistent with Prediction 2.

\subsection{Gas and Dark Matter Mass} 
From a theoretical perspective, we expect that classical Pop III star formation at $z \approx 6.6$ is likely to occur in two possible types of dark matter halos, depending on whether or not the region of the IGM they form in has been previously reionized. If the halo environment has not yet been ionized, LW feedback sets the relevant mass scale, and Pop III stars are expected to form in atomic cooling halos with virial temperatures of $T_{\rm vir}\approx 10^4~{\rm K}$ corresponding to virial masses of $M_{\rm h}\approx 5 \times 10^7~M_\odot$. This is because, by this late redshift, the LW background is high enough that Pop III star formation does not occur until halos are near the atomic cooling limit \citep[e.g., ][]{2020ApJ...897...95V}. In regions that have been ionized, we expect Pop III star formation to be delayed until halos are $M_{\rm h}\approx 3 \times 10^8~M_\odot$ due to photoheating of the gas \citep[e.g., ][]{2004ApJ...601..666D}. We examine the central gas content in these types of halos in hydrodynamical cosmological simulations. 

For halos formed in a neutral portion of the IGM, we consider the simulations of \cite{2019ApJ...885..127S}, that simulate three atomic cooling halos that undergo runaway collapse at $z\sim 11-13$ for a LW background of $J_{\rm LW}=10~J_{21}$, where $J_{21} \equiv 10^{-21}~{\rm erg~s^{-1}~cm^{-2}~Hz^{-1}~sr^{-1}}$. Within 3 Myr after collapse, the enclosed gas mass in the central 20 pc is ${\approx}10^5~M_\odot$ for the three halos.
We also have an example of a pristine halo that forms in a region of the IGM that has already been ionized in ``Halo C'' of \cite{2019ApJ...882..178K} \citep[and resimulated in][]{2025arXiv250112986S}. The enclosed gas mass within the central 20 pc of this halo is ${\sim}2\times 10^5~M_\odot$ (private communication). There is agreement with theoretical predictions and the more empirical estimates of the central gas mass from the LAP1-B observations \citep{LAP1B} (i.e., there is a central gas mass of ${\sim}10^5~M_\odot$).

Moving on to the dark matter component in LAP1-B, \cite{LAP1B} argue that the width of the H$\alpha$ line (${\sim}58.3 \pm 17.8~{\rm km~s^{-1}}$) suggests the presence of $\sim 10^7~M_\odot$ of dark matter within the central 20 pc. Simulations predict a substantially lower dark matter content within this radius in the halos we expect to host classical Pop III star formation. For instance, for the preionized case from \cite{2019ApJ...882..178K} and \cite{2025arXiv250112986S}, the central 20 pc contains equal amounts in dark matter and gas ($\sim10^5~M_\odot$ in each component).
Indeed, we note that, for an NFW profile with the expected concentration parameter of $c \approx 4$ \citep{2015MNRAS.452.1217C}, the enclosed mass within 20 pc is ${\sim}10^5~M_\odot$ for halos with $T_{\rm vir}\sim 10^4 ~{\rm K}$. 

We argue that the H$\alpha$ line broadening is not due to the orbital motions of the gas, but is instead a result of gas outflows. The observed line broadening could be due to gas motions as a result of photoionization, stellar winds, or supernovae winds from Pop III stars. We note that HII regions in local dwarf galaxies have line widths ranging from ${\sim}10-50~{\rm km~s^{-1}}$ \citep[][]{2022ApJ...929...74C}, the upper end of which is consistent with LAP1-B to within 1$\sigma$. There is some reason to expect Pop III systems to have particularly wide H$\alpha$ widths -- for example, the simulations of \cite{2004ApJ...610...14W} find radial velocities of $\sim 35 {\rm km~s^{-1}}$ in the HII regions around Pop III stars, which would be consistent with LAP1-B when summing over all directions. Pop III supernovae can drive gas to even higher velocities. In the 1D simulations of \cite{2005ApJ...630..675K}, a $10^{53}~{\rm erg}$ supernovae explodes within a $10^7~M_\odot$ dark matter halo at $z=20$ (i.e., the same virial temperature as the atomic cooling halo that could host LAP1-B at $z=6.6$). The initial outflow velocity is 1000's of ${\rm km~s^{-1}}$, but slows down to near zero within $20~{\rm pc}$ over 2.2 Myr.  Thus, the H$\alpha$ line width observed in LAP1-B could be the result of a supernovae as it decelerates within the first few Myr after the explosion. 
Therefore, we conclude that the observations are consistent with LAP1-B being in an atomic cooling halo, consistent with Prediction 1. We emphasize that neither of these simulations precisely matches the scenario of Pop III stars in an atomic cooling halo.  \cite{2004ApJ...610...14W} simulates a single Pop III star in a minihalo and \cite{2005ApJ...630..675K} is for a pair instability supernovae, which may be more energetic than the Pop III supernovae expected from lower mass Pop III stars. This motivates future hydrodynamical simulations of Pop III clusters in atomic cooling halos and their transition to metal-enriched star formation, but is beyond the scope of this work.

\subsection{Metal Enrichment}
We now consider the oxygen abundance of LAP1-B. We begin with the case where metals are primarily sourced by supernovae and then consider a case where they arise from stellar winds driven by rapidly rotating stars. 
Assuming that there is ${\sim}3\times 10^5~M_\odot$ of gas in the central region of LAP1-B as discussed above, the oxygen-to-hydrogen abundance ratio of $12+\log(\frac{\rm O}{\rm H}) = 6.31$ from \cite{LAP1B} implies that there is ${\sim}10~M_\odot$ of oxygen mixed in this region.
For the supernovae case, we assume that a small number of the Pop III stars have undergone supernovae and ejected their metals. 
According to \cite{2013ARA&A..51..457N}, the oxygen yield of one $40~M_\odot$ Pop III star is $8~M_\odot$. Thus, approximately one Pop III supernova is sufficient to explain the LAP1-B oxygen-to-hydrogen abundance ratio.
The precise assumption of $40~M_\odot$ stars is not required. For example, with the assumed IMF from \cite{LAP1B} ($1-100~M_\odot$), the oxygen yield per mass of stars is similar (a factor of two lower) than the hypothetical case of purely 40 $M_\odot$ stars.

It is also possible that some or all of the metals in LAP1-B were provided by stellar winds of rapidly rotating Pop III stars. The theoretical yield models of \cite{2012A&A...542A.113Y} predict that a $30~M_\odot$ star with an initial rotational velocity of 0.6 the Keplerian value yields $1.4~M_\odot$ of oxygen and a $150~M_\odot$ star with this initial velocity yields $3.23~M_\odot$. 
Thus, 7  $30~M_\odot$ or 3 $150~M_\odot$ rapidly rotating stars would provide the necessary 10~$M_\odot$ of oxygen.
That the O/H ratio matches with theoretical expectations from atomic cooling (or slightly larger reionized) halos means that LAP1-B is consistent with Predictions 1-3 described above.

Large theoretical uncertainties make the carbon-to-oxygen ratio of Pop III sources much more difficult to predict. For example, in the Pop III core collapse supernovae models from \cite{2010ApJ...724..341H}, the values of $\log{\rm (C/O)}$ vary over four orders of magnitude depending on uncertain parameters associated with the supernova explosion (e.g., energy). However, we point out that the value of $\log({\rm C/O}) \sim 0$ observed in LAP1-B is consistent with expectations for Pop III stars. For instance, the stars could be a mixture of ${\sim}40~M_\odot$ stars, some of which are rapidly rotating and produce winds with $\log({\rm C/O})=0.3$ \citep[see table 3 in][]{2023MNRAS.526.4467J} and some are not rapidly rotating and go supernovae with yields of $\log({\rm C/O})=-1.2$ \citep{2013ARA&A..51..457N}. We also note that the rapidly rotating stars of $150~M_\odot$ with initial velocity of 0.6 the Keplerian value at the equatorial surface from \cite{2012A&A...542A.113Y} have $\log({\rm C/O})=-0.13$, consistent with LAP1-B. Given the uncertainties in the supernovae yields form \cite{2010ApJ...724..341H}, there are portions of parameter space (e.g., supernovae explosion energy), which lead to C/O consistent with LAP1-B as well.

\section{Predicted Abundance of LAP1-B like objects}
\label{sec:abundance}
Next, we turn our attention to predicting the abundance of classical Pop III sources to see if the discovery of LAP1-B is expected. We use the semi-analytic model originally developed in \cite{2020ApJ...897...95V} with the modifications described in \cite{2025arXiv250614482V} (the most significant of which is calibrating the critical halo mass for Pop III star formation from \cite{2021ApJ...917...40K}).
The foundation of the model is 10 N-body simulation boxes from \cite{2020ApJ...897...95V}, which are each 3 Mpc across and contain $512^3$ particles (giving a particle mass resolution of $8,000~M_\odot$). We ran these simulations with {\sc gadget2} \cite{2001NewA....6...79S} and generated halo merger trees with {\sc rockstar} with {\sc consistent trees} \cite{2013ApJ...762..109B, 2013ApJ...763...18B}. {\sc rockstar} has been shown to accurately measure halo masses and positions with $\gtrsim 20$ particles. This corresponds to a halo mass of $M_{\rm h}\approx 2\times 10^5 ~ M_\odot$ for our runs (which is sufficient to resolve halos hosting Pop III star formation).
The semianalytic model includes a variety of important feedback processes including LW radiation, external metal enrichment from supernovae winds, and cosmic reionization. Each of these processes are followed including the 3-dimensional positions of halos and the resulting ioniziation/metal bubbles. We also note that the model includes baryon-dark matter streaming velocities that suppress star formation at high redshift in low-mass halos \citep{2010PhRvD..82h3520T,2012MNRAS.424.1335F}. For more details on the model, see \cite{2020ApJ...897...95V, 2025arXiv250614482V}.

We first compute the abundance of Pop III sources from our simulation boxes. This is done by counting the number of new Pop III-forming halos at each simulation snapshot which are spaced in time by 1/40 times the age of the Universe (which is ${\sim}20~{\rm Myr}$ at $z=6.6$) and converting to a number density using the streaming-velocity weighting in Eq.~2.1 of \cite{2025arXiv250614482V}. Additionally, the number density is multiplied by a factor that accounts for the duration of time the Pop III cluster will appear like LAP1-B. This factor is given by $\Delta t_{\rm vis}/\Delta t_{\rm snap}$, where $\Delta t_{\rm snap}$ is the time between simulation snapshots and $\Delta t_{\rm vis}$ is the duration of time the source will appear similar to LAP1-B, which we assume is 3 Myr since this is roughly the lifetime of massive Pop III stars. Thus, operationally we estimate the number density of visible Pop III halos by finding the number density of new Pop III sources formed between snapshots and multiplying by our weighting factor (0.15 at $z{\sim}6.6$). This estimate assumes that Pop III stars form uniformly in time between snapshots, which is a good approximation for the duration of time between our snapshots.
We assume the same fiducial model parameters from \cite{2025arXiv250614482V} and find that all of the Pop III sources occur in neutral regions of the IGM, and thus at $z\approx 6.6$ are in $T_{\rm vir}{\sim}10^4~{\rm K}$ halos due to LW feedback. 

\begin{figure*}
    \centering
    \includegraphics[width=17cm]{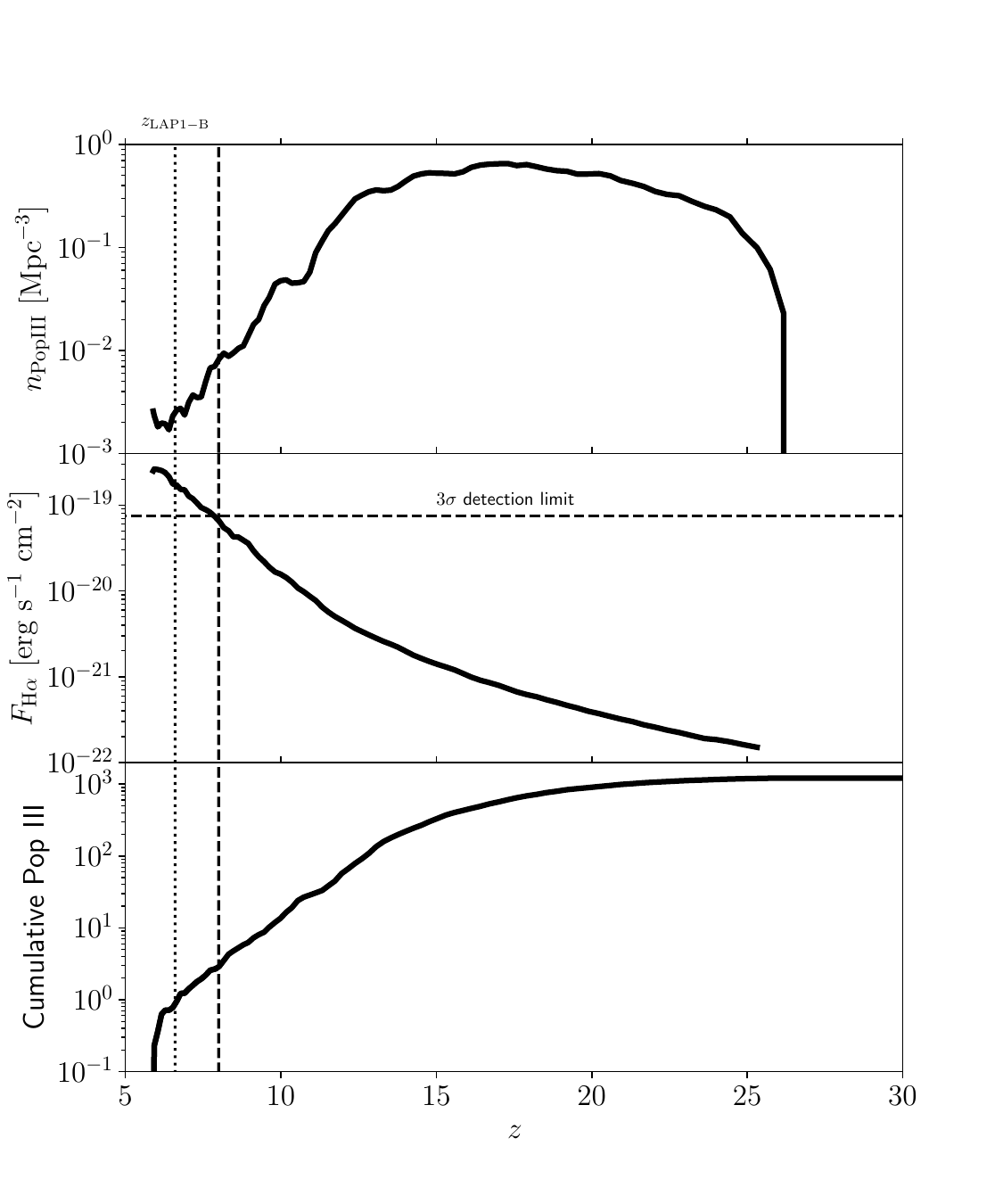}
    \caption{ \emph{Top panel}: Abundance of classical Pop III sources that have formed within the past 3 Myr. \emph{Middle Panel}: The mean H$\alpha$ flux from these sources assuming a magnification of $\mu=100$. The horizontal dashed line corresponds to the 3$\sigma$ uncertainty in \cite{LAP1B}. \emph{Bottom Panel}: The predicted cumulative number of Pop III sources formed within 3 Myr at redshift from 5.9 to $z$ within the area where galaxy cluster MACS J0416 has magnification $\mu>30$. The dashed vertical line corresponds to the redshift of $z=8$ where the detection of the H$\alpha$ line is at ${\sim}3\sigma$. The dotted vertical line indicates the redshift of LAP1-B, $z=6.6$. The cumulative abundance at $z=6.6$ is 0.88, consistent with the detection of LAP1-B.
    \label{fig:abundance}  }
\end{figure*}

In the top panel of Figure \ref{fig:abundance}, we present the number density of active Pop III sources as a function of redshift (i.e., the number within 3 Myr of the start of star formation). We also show the corresponding mean H$\alpha$ flux that would be observed from these Pop III sources in the middle panel. 
The mean Pop stellar mass of a newly formed Pop III cluster at $z=6.6$ is $5000~M_\odot$ in our fiducial model. Thus, to match the H$\alpha$ flux with \cite{LAP1B} we have assumed only 30 percent of this is currently in active Pop III stars. This is equivalent to applying a modification of 0.3 to the highly uncertain Pop III star formation efficiency parameter in our model. We also assume an escape fraction of 10 percent even though the semi-analytic model uses a value of 50 percent for all Pop III halos. Simulations find that atomic cooling halos have lower escape fractions than minihalos \citep[e.g., ][]{2016ApJ...833...84X}. Explicitly modifying the semi-analytic model to account for this would not significantly change our results because reionization feedback is dominated by metal-enriched sources by the lower redshifts when atomic cooling halos typically host Pop III star formation. We note that the standard deviation in the H$\alpha$ flux at fixed redshift is roughly 10 percent of the mean value for $z\lesssim  8$ because the halo masses where Pop III star formation occurs are more sensitive to the global background than local fluctuations in the LW flux and we assume a constant star formation efficiency. The reduction in flux at higher redshifts is a combination of increased cosmological luminosity distance and the fact that Pop III star formation occurs in lower mass halos with fewer stars as a result of the lower LW flux at earlier times.

We can also use these results to compute the number of newly formed sources that appear in regions where the MACS J0416 galaxy cluster has a lensing magnification of $\mu>30$. This corresponds to an area of $0.2~{\rm arcmin}^2$ on the sky \citep[see Table 6 in][]{2015ApJ...800...38G}. In the bottom panel of Figure \ref{fig:abundance}, we present the cumulative number of Pop III sources behind MACS J0416 at this magnification, along with the redshift at which their mean H$\alpha$ line flux drops below a $3\sigma$ detection. We assume no Pop III sources at $z<5.9$. This is where our N-body simulations end, but also given that we find all of the Pop III sources in non-ionized atomic cooling halos, the number density is likely to drop significantly after reionization. We clearly see from Figure \ref{fig:abundance} that finding a Pop III galaxy near $z=6.6$ is likely, while the numerous sources at high redshift drop below the flux limit. This shows that an initial detection of Pop III stars by \emph{JWST} targeting clustering magnification is expected to find a source at $z\sim 6.5$, consistent with LAP1-B. Given that the number density of dark matter halos is well understood in the context of the standard model of cosmology, the agreement between our abundance estimate and the observation of LAP1-B provides strong support that LAP1-B is hosted by an atomic cooling halo, in agreement with Prediction 1 introduced in Section \ref{sec:intro}.

Finally, we note that LAP1-B is detected within ${\sim}300~{\rm pc}$ of an additional fainter source, LAP1-A \citep{2023A&A...678A.173V}. The virial radius of an atomic cooling halo at $z\sim6.6$ is ${\sim}1500~{\rm pc}$. This suggests that LAP1-A and LAP1-B may both reside within the same dark matter halo (depending on their unknown separation along the line-of-sight). This could be the result of a recent merger between an atomic cooling halo hosting LAP1-B and a similar atomic halo hosting LAP1-A. Using the Sheth-Tormen halo mass function \citep{1999MNRAS.308..119S} and the two-point correlation function at $z\sim10$ from \cite{2014MNRAS.445.1056V}, we find that there is a ${\sim}$20 percent chance an atomic cooling halo will have a subhalo that is within a factor of a few of the same halo mass. This indicates that it is quite common for halos near the atomic cooling limit to have recently undergone a major merger and not yet mixed into a single distinct halo. LAP1-A is much fainter in H$\alpha$ than LAP1-B even though it is thought to be gravitationally magnified by a factor of 5 or more stronger than LAP1-B \citep{2023A&A...678A.173V}. Thus, if LAP1-A is a similar system, it either formed fewer Pop III stars or those stars have already died. However, it is difficult to interpret LAP1-A confidently  without deeper observations or detailed hydrodynamical simulations of such a merger scenario.

\section{Discussion and Conclusions}
\label{sec:conclusions}
We argue that LAP1-B \citep{2023A&A...678A.173V,LAP1B} is the first Pop III candidate to agree with three key theoretical predictions for classical Pop III sources (i.e., not mixtures of Pop III and metal-enriched stars in evolved galaxies). The first prediction (Prediction 1) is that Pop III stars form in halos with $T_{\rm vir}\sim 10^3- 10^4~{\rm K}$. Our fiducial semi-analytic model \citep[unaltered from previous work][]{2025arXiv250614482V}  finds that Pop III stars form in $T_{\rm vir}{\sim}10^4~{\rm K}$ dark matter halos with a number density that is consistent with the discovery of LAP1-B, suggesting that it is hosted in such a halo. This halo mass is also consistent with the observed oxygen-to-hydrogen abundance ratio, given the amount of central gas predicted by hydrodynamical cosmological simulations and the stellar mass in Pop III stars estimated from the H$\alpha$ line luminosity. This halo mass implies that the velocity width of the H$\alpha$ line in LAP1-B is not due to ${\sim} 10^7~M_\odot$ of dark matter claimed by \cite{LAP1B} (which would require a much more massive halo), but is instead due to outflows caused by photoionization, supernovae, and/or stellar winds. The velocities of the theoretical predictions for these outflows are consistent with the observed linewidth. The second prediction (Prediction 2) is that Pop III stars have a logarithmically flat IMF (likely over a massive range such as $10-1000~{M_\odot}$). This is consistent with the CIV1549-to-[OIII]5007  line ratio as described in \cite{LAP1B} and shown in their Fig.~3. It is also consistent with the abundances of oxygen and carbon measured by \cite{LAP1B} as discussed in Section \ref{sec:properties}. The third prediction (Prediction 3) is that Pop III stars form in low-mass clusters of total stellar of ${\sim}1000$ M$_{\odot}$, before being polluted by metals and transitioning to metal-enriched star formation. The line luminosity of H$\alpha$ closely matches this stellar mass.

Several other \emph{JWST} Pop III candidates have recently been reported \citep{2024ApJ...967L..42W, 2024A&A...687A..67M,glimpse, 2025arXiv250710521M, 2025arXiv250717820C}. However, the interpretation of all of these candidates requires Pop III stellar masses of $\gtrsim 10^{5-6}~M_\odot$. 
Such high stellar masses are not predicted in the case of classical Pop III star formation. We note that in the simulations of \cite{2022ApJ...935..174S}, which include a subgrid model for turbulent mixing of metals in gas, they find that it takes tens of Myrs for metals to mix uniformly through the gas. This leads to mixtures of Pop III and metal-enriched stars and a substantial increase in the overall abundance of Pop III stars. However, in a representative  halo at $z=8$, they find a total mass of ${\sim}5\times 10^4~M_\odot$ of Pop III stars ever produced and only ${\sim}3\times 10^3~M_\odot$ which are alive at this time, substantially less than needed to explain previous \emph{JWST} Pop III candidates.

We conclude by emphasizing that our semi-analytic model shows that, for the sensitivity of the observations from \cite{LAP1B}, the most likely redshift for a Pop III observation is $z\approx 6.5$. However, observations with increased sensitivity are predicted to find additional sources skewed towards higher redshifts (e.g., $>10$ by $z=10$ with observations an order of magnitude more sensitive compared to \cite{LAP1B}). Thus, LAP1-B may only represent the tip of the iceberg in terms of the study of Pop III stars with gravitational lensing from galaxy clusters. Finally, while here we have emphasized the consistenncy between predictions and observations, future exploration of the parameter space in our semi-analytic model to determine where it is compatible with observations, may constrain the properties of Pop III star formation (e.g., Pop III star formation efficiency and delay time between Pop III and metal-enriched star formation).

\begin{acknowledgments}
EV and RH are supported by NSF grant AST-2009309 and NASA ATP grant 80NSSC22K0629. All runs of the semi-analytic model were performed at the Ohio Supercomputer Center. 
G.L.B. acknowledges support from the NSF (AST-2108470 and AST-2307419, ACCESS), a NASA TCAN award, and the Simons Foundation through the Learning the Universe Collaboration.
\end{acknowledgments}

\bibliography{LAP1B}{}
\bibliographystyle{aasjournalv7}



\end{document}